\documentclass[final,3p]{elsarticle}
\usepackage{amsmath}
\usepackage{amssymb}
\usepackage{graphicx}
\usepackage{tikz}
\usepackage{comment}
\usepackage{tabularx}
\usepackage{booktabs}
\usepackage{array}
\usepackage{float}
\usepackage{algorithm} 
\usepackage{algpseudocode} 
\usepackage{xcolor,graphicx,subfig,lipsum}
\usepackage{enumitem}
\usepackage{multirow}
\usepackage{hyperref}
\usepackage{comment}

\begin{document}

\begin{frontmatter}

\title{A Lightweight Deep Learning-based Model for Ranking       Influential Nodes in Complex Networks}
\author[inst1]{Mohammed A. Ramadhan\texorpdfstring{\corref{cor1}}{}}
\ead{mohammed.abdullah@staff.uoz.edu.krd}

\author[inst1,inst2]{Abdulhakeem O. Mohammed}
\ead{a.mohammed@uoz.edu.krd}

\affiliation[inst1]{organization={Computer Science Department, College of Science, University of Zakho}, 
            addressline={Duhok 42002}, 
            city={Kurdistan Region}, 
            country={Iraq}}
\affiliation[inst2]{organization={Department of Computer Science and Information Technology, The American University of Kurdistan},
            country={Kurdistan Region, Iraq}}

\cortext[cor1]{Corresponding author}

\begin{abstract}
Identifying influential nodes in complex networks is a critical task with a wide range of applications across different domains. However, existing approaches often face trade-offs between accuracy and computational efficiency. To address these challenges, we propose 1D-CGS, a lightweight and effective hybrid model that integrates the speed of one-dimensional convolutional neural networks (1D-CNN) with the topological representation power of GraphSAGE for efficient node ranking.
The model uses a lightweight input representation built on two straightforward and significant topological features: node degree and average neighbor degree. These features are processed through 1D convolutions to extract local patterns, followed by GraphSAGE layers to aggregate neighborhood information.
We formulate the node ranking task as a regression problem and use the Susceptible–Infected–Recovered (SIR) model to generate ground truth influence scores. 
1D-CGS is initially trained on synthetic networks generated by the Barabási–Albert model and then applied to real world networks for identifying influential nodes.
Experimental evaluations on twelve real world networks demonstrate that 1D-CGS significantly outperforms traditional centrality measures and recent deep learning models in ranking accuracy, while operating in very fast runtime.
The proposed model achieves an average improvement of 4.73\% in Kendall’s Tau correlation and 7.67\% in Jaccard Similarity over the best performing deep learning baselines. It also achieves an average Monotonicity Index (MI) score 0.99 and produces near perfect rank distributions, indicating highly unique and discriminative rankings. Furthermore, all experiments confirm that 1D-CGS operates in a highly reasonable time, running significantly faster than existing deep learning methods, making it suitable for large scale applications.

\end{abstract}

\end{frontmatter}

\section{Introduction}\label{sec:Introduction}
The ability to accurately identify influential nodes in complex networks is essential for a wide range of applications, including epidemic control \cite{Epdmic-control-ye2023vital}, viral marketing \cite{marketing-2009discovering}, social networks \cite{Social1-2011small,social2-2021golden,social3-2018top,socialnetwork-ishfaq2022identifying}, etc. These influential nodes are those whose position in the network allows them to play a critical role in spreading information or influence to a large portion of the network. However, the challenge lies not only in achieving high accuracy in identifying these nodes, but also in doing so efficiently, especially as real world networks grow in size and complexity.

In recent years, there has been significant research on identifying influential nodes in complex networks. Centrality based methods are largely based on the concept of centrality, which quantifies a node's influentiality based on its topological position within the network. These methods can be broadly categorized into four groups \cite{lu2016vital}: (a) Neighborhood-based centrality, such as Degree Centrality (DC) \cite{DCfreeman-FREEMAN1978215}, which considers only the number of immediate connections a node has;
(b) Path-based centrality, including Closeness Centrality (CC) \cite{CC1966centrality} and Betweenness Centrality (BC) \cite{BC-freeman1977set}, which evaluate node influentiality by analyzing the shortest paths or information flow through the network;
(c) Iterative-based centrality, where node rankings are refined through repeated computations, as in Eigenvector Centrality (EC) \cite{EC2007}, PageRank (PR) \cite{pagerank1998}, Cluster Rank (CR) \cite{CR-2013identifying}, Leader Rank (LR) \cite{LR-2011leaders}, and VoteRank (VR) \cite{VR2016identifying};
and (d) Node operation-based centrality, which involves manipulating or analyzing node influence through removal or subgraph-based strategies, such as K-core \cite{k-shell2010identification} and the K-shell Iteration Factor (KSIF).
Some of these methods, especially path-based and iterative approaches, are computationally expensive and scale poorly to large networks.
Others, such as neighborhood-based centrality, are computationally efficient and interpretable, but they often fail to capture the complex relationships and global dynamics present in real world networks. 

In response to these limitations, Machine Learning (ML) and Deep Learning (DL) based methods have emerged as promising alternatives. Most ML-based methods rely on feature engineering to construct descriptive feature vectors for each node, typically incorporating centrality measures, neighborhood statistics, or topological roles, which are then fed into traditional learning models to classify or rank influential nodes \cite{ls-svm-wen2018lsvm,rezaei2023machine, zhao2020-7machine, ML-and-Centralityhajarathaiah2024node}. While these approaches offer greater flexibility and adaptability than centrality based methods, they often lack the ability to capture deeper topological patterns present in complex networks.

On the other hand, DL methods, especially those based on 2D-Convolutional Neural Network (2D-CNN) and Graph Neural Network (GNN), aim to learn topological representations directly from network structure. Examples of these models, such as CGNN \cite{Anew-approach-zhang2022new}, infGCN \cite{infGCN-zhao2020}, RCNN \cite{yu2020rcnn}, and MRCNN \cite{ou2022iM-RCNN} have demonstrated strong performance in extracting complex patterns and accurately predicting node influence. However, despite their effectiveness, these methods often come with significant computational overhead, particularly in large scale networks. This is largely due to deep architectural designs, costly message passing operations, and slow convergence during training. As a result, such models can be difficult to implement in time critical scenarios, where both accuracy and efficiency are crucial.
To overcome these challenges, we propose a new hybrid deep learning model, named 1D-CGS, that delivers both high ranking accuracy and low computational cost. Our method combines the strengths of two efficient and scalable components: a 1D-Convolutional Neural Network (1D-CNN) that captures  local topological relationships in node level features and a GraphSAGE \cite{GraphSage2017inductive} model that learns context aware embeddings through neighborhood aggregation.
This lightweight yet powerful combination enables our model to learn both local and topological patterns without the heavy computational cost associated with deeper GNN architectures.

We evaluated our model on 12 real world networks of varying sizes and domains. Our experimental results demonstrate that 1D-CGS consistently outperforms centrality-based methods and existing deep learning methods in terms of ranking accuracy, while achieving significantly faster runtime on real world network datasets.
The main contributions of this work are as follows:
\begin{itemize}

\item We introduce a hybrid model that integrates 1D-CNN with GraphSAGE to effectively capture both local node features and global topological patterns within a lightweight and scalable design.

\item We construct a minimal yet informative input representation based on degree and average neighbor degree, which are fast to compute and effective for influence prediction.

\item The proposed model significantly reduces training and testing time compared to existing deep learning models such as RCNN and MRCNN, as shown by extensive experiments on synthetic and real world networks.

\item The experimental results demonstrate that the 1D-CGS model consistently outperforms the second best deep learning-based method across multiple real world networks in ranking influential nodes. On average, 1D-CGS achieves a 4.7275\% improvement in Kendall’s Tau Correlation and a 7.67\% improvement in Jaccard Similarity, indicating more accurate and consistent ranking performance.

\end{itemize} 

The rest of the paper is organized as follows: Section \ref{sec:relatedworks} reviews related work,
Section \ref{sec:Method} presents the proposed methodology and materials, Section \ref{sec:experement and discussion} presents experiments and discussion, and Section \ref{sec:conclusion} concludes the study.  
The source code and experimental network datasets used in this study are publicly available at: \href{https://github.com/mohammed-ar96/1D-CGS}{https://github.com}.

\section{Related work} \label{sec:relatedworks}
Identifying influential nodes in complex networks has been extensively studied, with approaches broadly categorized into centrality-based, ML-based, and DL-based methods. In this section, we review these approaches and their limitations.

Many studies have focused on developing centrality-based methods that rely on specific aspects of network topology. However, these methods such as BC \cite{BC-freeman1977set}, CC \cite{CC1966centrality}, DC \cite{DCfreeman-FREEMAN1978215}, EC \cite{EC2007}, and PR \cite{pagerank1998} often suffer from limitations in both performance and flexibility, as they typically consider only a single topological perspective of the network.

%A lot of work has been done to design centrality-based methods that rely on specific parts of the network's topological structure. However, these methods often suffer from limitations in performance and flexibility. For example, betweenness centrality [17] and closeness centrality [18] focus on global shortest paths, while degree centrality [19] and semi-local centrality [20] consider only local neighborhood connections. Other methods, such as eigenvector centrality [21] and PageRank [22], introduce iterative computations to capture the importance of neighbors. Despite their widespread use, these approaches generally focus on a single structural perspective, which limits their ability to accurately identify influential nodes in real-world networks.

To address these issues, machine learning-based methods have emerged as an alternative. Research in this direction mainly focuses on feature engineering. For instance, Rezaei et al. \cite{rezaei2023machine} reformulated the problem of identifying influential nodes as a regression task. They performed collective feature engineering, constructing feature vectors for each node using attributes such as connectivity, degree, and extended coreness, and then applied machine learning models to rank the nodes accordingly. Similarly, in \cite{zhao2020-7machine, ML-and-Centralityhajarathaiah2024node} the problem was treated as a classification task. Each node was represented by a feature vector built using centrality-based method like DC, BC, CC, and other measures, which were then used to train ML models to identify influential nodes. While machine learning methods provide more flexibility and adaptability than traditional centrality-based methods, they still often struggle to capture deeper topological dependencies in the network.

On the other hand, deep learning-based methods, especially those based on 2D-CNN and Graph Neural Networks (GNNs), aim to learn topological patterns directly from the network data. For example, a study by Yu et al. \cite{yu2020rcnn} reformulated the influence ranking task as a regression problem. They created an adjacency matrix for each subgraph, transformed it based on the degree of the node, and fed it into a CNN where it was processed using 2D convolutional layers. In a similar direction, Ou et al. \cite{ou2022iM-RCNN} extended the RCNN model by proposing MRCNN, which used the same 2D-CNN structure but with three input channels instead of one, the input was based on V-community, neighborhood degree, and k-core values of the node.
Another relevant work by Zhang et al. \cite{Anew-approach-zhang2022new}, where they proposed CGNN, a hybrid model combining 2D-CNN with GNN components. They constructed subgraphs for each node, applied a contraction algorithm to generate matrices, and then fed these into the hybrid CNN and GNN model. While these deep learning models achieve better performance than traditional ML-based methods mainly due to their deeper architectures and ability to extract complex features they often require significant computation time and resources.

In our work, we aim to reduce this overhead by constructing a fast and effective input representation based on degree and average neighborhood degree of each node. To keep the model lightweight, we use 1D-CNN instead of 2D-CNN, and we enhance the topological learning by integrating GraphSAGE, which helps aggregate neighborhood information efficiently. This design proved to be both fast and powerful, delivering very competitive results in terms of ranking accuracy, while significantly reducing training and testing time compared to previous studies (see Sections \ref{sec:results} and \ref{sec:RT}).

\section{Method} \label{sec:Method}
\begin{comment}
    In this section, we present the architecture and implementation details of the 1D-CGS model. The model integrates 1D-CNN with GraphSAGE in a sequential manner. It first applies convolutional operations to transform raw node features and then aggregates structural information through neighborhood sampling and message passing. This design balances efficient feature representation with scalable structural learning. An overview of the model architecture is provided in Figure 1.
\end{comment}

\begin{figure}[H] 
    \centering
    \includegraphics[width=\textwidth]{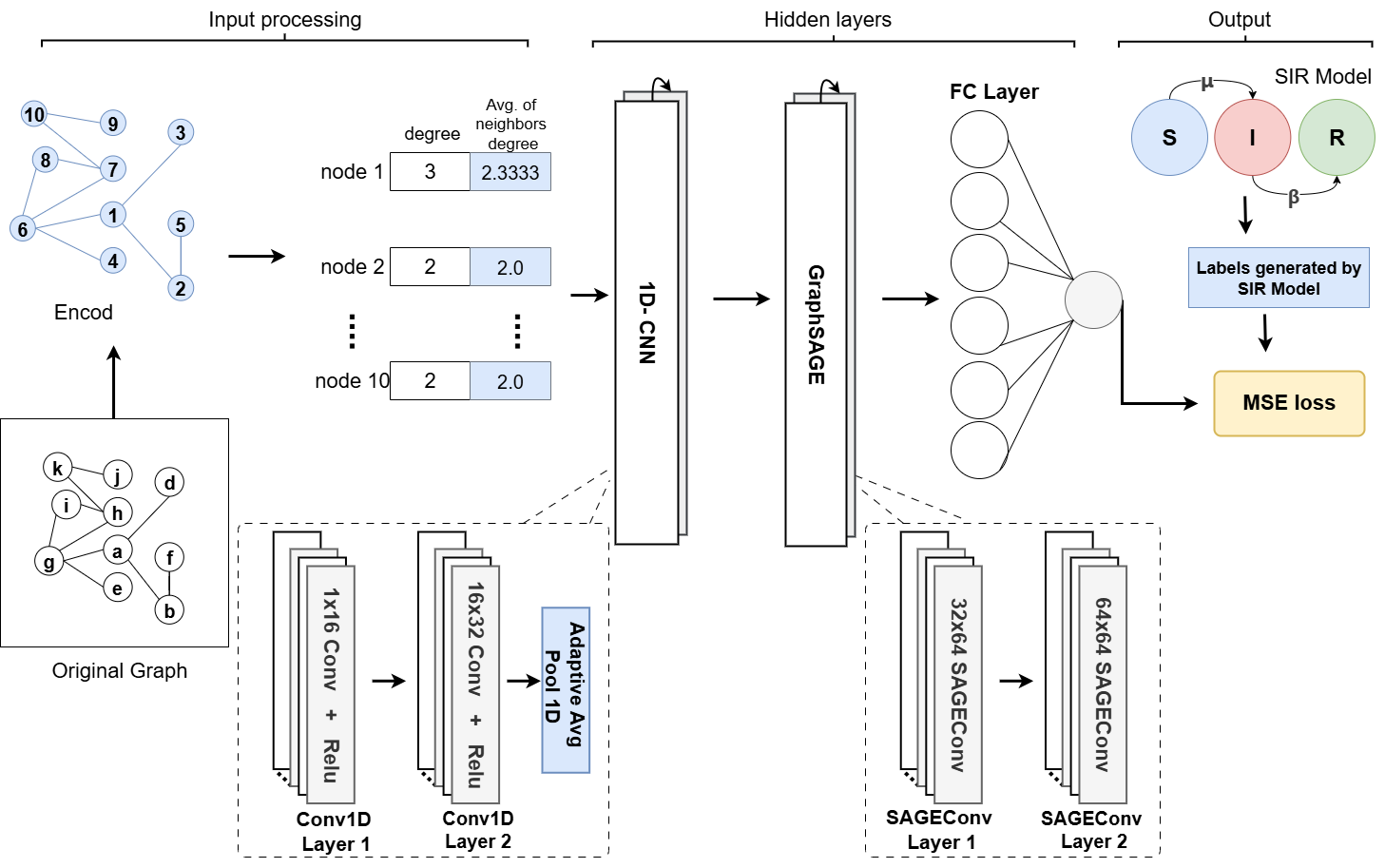} % Direct width control
    \caption{The 1D-CGS model computes node influence by combining local feature extraction and topological aggregation. Each node is represented by a compact 2-dimensional feature vector, degree and average neighbor degree processed by a 1D-CNN. The resulting embeddings are then passed through two GraphSAGE layers to capture relational dependencies from the network structure. Finally, a fully connected layer maps the aggregated features to a scalar influence score.}
    \label{img1:methodology}
\end{figure}

In this section, we present the architecture and implementation details of 1D-CGS model. We begin by constructing feature vector of each node using: the degree and the average neighbor degree of each node in the network. Next, we describe the process of generating node influence labels through simulations of the SIR \cite{SIR2001epidemic} model, which provide realistic approximations of node influence. Finally, we detail the architecture of the 1D-CGS model, including its input, hidden, and output layers. An overview of the model architecture is provided in Figure \ref{img1:methodology}.

\subsection{Feature Construction} \label{sec:feature constructio}
To represent each node in the network, we construct a feature vector composed of two local topological descriptors: degree and average neighbor degree. These features are selected to ensure that the model remains computationally efficient and scalable, particularly when applied to large networks. Both features are simple to compute and require only local information, avoiding the high cost associated with more complex topological metrics. Let $G=(V,E)$ be an undirected, unweighted graph, where $V$ is the set of nodes with $|V|=n$ and $E$ is the set of edges. For each node $v_i \in V$, the following features are extracted:
\newline \newline
\textbf{Node degree.} The degree of a node, is the number of immediate neighbors it has. This is one of the most basic measures and serves as a simple indicator of potential influence. It is defined as $\deg(v_i)$.
\newline \newline
\textbf{Average Neighbor Degree (AND).} This feature represents the average degree of all nodes directly connected to $v_i$. It captures the connectivity strength of a node’s neighborhood and is computed as:
\begin{equation}
\mathrm{AND}(v_i) = \frac{1}{|N(v_i)|} \sum_{v_j \in N(v_i)} \deg(v_j)
\end{equation}
where $N(v_i)$ is the set of nodes directly connected to $v_i$, $|N(v_i)|$ is the number of neighbors, and $\deg(v_j)$ is the degree of the neighbor node $N(v_j)$. This feature captures the connectivity level of the local neighborhood around each node and complements the raw degree information by considering not just how many connections a node has, but also how influential its neighbors are.

Each node in the network is represented by a feature vector $x_i$ which serves as input to the 1D-CGS model.
\[
x_i = \left[\deg(v_i), \mathrm{AND}(v_i)\right]
\]
\subsection{Generating Labels} \label{sec:generating labels}
As the 1D-CGS model is trained using a supervised learning model, it requires a numerical label for each node to guide the learning process. In this work, we formulate the task of identifying influential nodes as a regression problem, where the label corresponds to the estimated spreading influence of each node. we generate these labels through simulation. Specifically, we adopt the SIR model, a well established framework for modeling information or disease propagation in complex networks. 
In the SIR model, each node can be in one of three possible states: Susceptible (S), Infected (I), or Recovered (R). At the start of the simulation, a single node is initialized as the infected node, while all other nodes are set to the susceptible state. As the process unfolds: (a) each infected node attempts to infect its susceptible neighbors with a fixed infection probability 
$\mu$; (b) infected nodes transition to the recovered state with a recovery probability $\beta$; (c) the simulation continues until no nodes remain in the infected state.
The spreading influence of node $v_i$ is defined as the total number of nodes that become infected (and subsequently recovered) during the simulation when the spread originates from $v_i$ . Due to the stochastic nature of the SIR model, we repeat the simulation 1,000 times for each node and compute the average number of infected nodes across these trials. This average value is assigned as the node’s influence label for training. 

\subsection{Model} \label{sec:model}
The 1D-CGS model is a sequential hybrid architecture designed to learn both local and topological patterns from node features and network structure. It combines the ability of a 1D-CNN with the inductive learning power of GraphSAGE. The overall architecture consists of three main stages: (1) an input layer that processes node features, (2) a hidden representation module composed of a 1D-CNN followed by two GraphSAGE layers, and (3) an output layer that produces continuous valued influence scores.

\textbf{Input layer.} The model receives as input a feature matrix $X \in \mathbb{R}^{n \times 2}$, where each row $x_i$ corresponds to a node $v_i$ in the graph and contains two topological features: degree and average neighbor degree of the node, as described in Section \ref{sec:feature constructio}. These features are reshaped to $[n,1,2]$ in order to fit the input format expected by 1D convolutional layers,
$n$ is the total number of nodes.

\textbf{Hidden layers.} The feature matrix is first passed through two consecutive 1D convolutional layers with ReLU activation functions. Each layer uses a kernel size of 3 with padding of 1 and stride of 1. The first convolutional layer outputs 16 channels, and the second outputs 32. After convolution, an adaptive average pooling layer reduces the temporal dimension, followed by a flattening operation, resulting in a transformed feature representation $z_i \in \mathbb{R}^{32}$ for each node. Then the 32-dimensional vectors produced by the CNN are then used as input to a two layer GraphSAGE network. GraphSAGE updates each node’s embedding by aggregating features from its local neighborhood using a mean aggregator. Let $h_i^{(0)} = z_i$ be the initial input to GraphSAGE. The node embedding is updated layer by layer as follows:
\begin{equation}
h_i^{(l+1)} = \sigma\left(\mathbf{W}^{(l)} \cdot \mathrm{MEAN}\left(\{h_i^{(l)}\} \cup \{h_j^{(l)} : v_j \in N(v_i)\}\right)\right)
\end{equation}
where $h_i^{(l)} \in \mathbb{R}^d$ is the embedding of node $v_i$ at layer $l$, $\mathbf{W}^{(l)} \in \mathbb{R}^{d \times d}$ is a learnable weight matrix, $\sigma(\cdot)$ is the ReLU activation function, $N(v_i)$ is the set of neighboring nodes of $v_i$, and $\mathrm{MEAN}$ is the neighborhood aggregation function.

\textbf{Output layer.} The output layer consists of a single fully connected linear layer that maps the final 64-dimensional node embeddings to scalar values. For each node $N(v_i)$, the predicted influence score $\hat{y}_i$ is computed as:
\begin{equation}
\hat{y}_i = \mathbf{w}^\top h_i + b
\end{equation}
where $h_i \in \mathbb{R}^{64}$ is the final node embedding, $\mathbf{w} \in \mathbb{R}^{64}$ is the learnable weight vector, $b \in \mathbb{R}$ is the bias term, and $\hat{y}_i \in \mathbb{R}$ is the predicted influence score.
The model is trained using the mean squared error (MSE) loss function $\mathcal{L}$, which measures the discrepancy between the predicted influence scores $\hat{y}_i$ and the ground truth labels $y_i$ obtained from the SIR simulation:

\begin{equation}
\mathcal{L} = \frac{1}{n}\sum_{i=1}^n (y_i - \hat{y}_i)^2
\end{equation}
Training is performed using the Adam optimizer \cite{AdamOp-2014adam} with a learning rate of 0.005 for 3000 epochs. 
To assess the model's behavior under controlled conditions, 1D-CGS was trained on synthetic Barabási–Albert (BA) networks \cite{BAnet-2010catastrophic} of varying sizes and average degrees. Empirical results showed that the model achieved the most stable performance and fastest convergence on a BA network with $n=1000$ nodes and an average degree of $k=4$. This configuration was therefore used as the primary setting for training and benchmarking in our experiments, and building on prior work \cite{yu2020rcnn, ou2022iM-RCNN}, we simulated the SIR model (see Section \ref{sec:generating labels}) using $\mu= 1.5\mu_c$ as the infection rate, where $\mu_c$ is the SIR model spreading threshold, which defined as:
\begin{equation}
    \mu_c = \frac{\langle k \rangle}{\langle k^2 \rangle - \langle k \rangle}
\end{equation}
where $\langle k \rangle$ is the average degree of the network.

\section{Experiments and Discussion} \label{sec:experement and discussion}
To evaluate the effectiveness and efficiency of the proposed 1D-CGS model, we conducted extensive experiments on a diverse set of real world networks. This section presents the network datasets used, evaluation metrics, the baseline methods for comparison, and a comprehensive discussion of the results.

\subsection{Network Datasets} The 1D-CGS model trained on synthetic networks generated using the BA model. We explored several configurations by varying the number of nodes [1000, 2000, 3000, 4000, 8000] and the average degree [4, 10, 20]. Experimental results showed that the model achieves the most stable performance and fastest convergence when trained on a network with 1000 nodes and an average degree of 4, which is used as the default training setting throughout the study, as mentioned in Section \ref{sec:model}. For evaluation, the trained model was tested on 12 diverse real world networks from various domains, including Jazz \cite{jazz2003community}, Email \cite{email-guimera2003self}, Stelzl \cite{stelzl2005human}, Figeys \cite{figeys-2007large}, Hamster \cite{hamster-kunegis2013konect}, Facebook \cite{facebook2012learning}, EPA \cite{net-web}, Router \cite{router-spring2004measuring}, GrQ \cite{GrQ2007graph}, LastFM \cite{lastFM2020characteristic}, PGP \cite{PGP-boguna2004models}, and Sex \cite{Sex-2011simulated}. These networks differ in size, density, and structure, providing a comprehensive benchmark for assessing the model’s generalization capability. Table \ref{table:topology} summarizes key topological statistics of these networks, including the number of nodes, edges, density, average degree, maximum degree, and infection thresholds used in the SIR influence simulations.

\newcolumntype{C}{>{\centering\arraybackslash}X}
\begin{table}[H] 
\centering
\resizebox{\textwidth}{!}{%
\fontsize{8pt}{9pt}\selectfont
\begin{tabularx}{\textwidth}{lCCCCCC}
\toprule
\textbf{Networks}	& \textbf{Nodes}	& \textbf{Edges} & \textbf{Density}& \textbf{${\langle K \rangle}$} & \textbf{$K_{max}$} & 
 \textbf{$\mu_c$}\\
\midrule
Jazz	& 198 & 2742  & 0.1406 & 27.697 & 100 & 0.0265 \\
Email	& 1133	& 4351  & 0.0085 & 9.622 & 71 & 0.0565  \\
Stelzl	& 1706	 & 3155  & 0.0022 & 3.699 & 95 & 0.0632 \\
Figeys & 2239 & 6432  & 0.0026 & 5.745 & 314 & 0.0181 \\
Hamster	& 2426 & 16630  & 0.0057 & 13.711 & 273 & 0.0240 \\
Facebook & 4039	& 88234 & 0.0108 & 43.691  & 1045 & 0.0094 \\
EPA		& 4271	& 8909  & 0.0010 & 4.172 & 175 & 0.0366\\
Router	& 5022	& 6258  & 0.0005 & 2.492   & 106 & 0.0786 \\
GRQ	& 5242 & 14484  & 0.0011 & 5.526 & 81 & 0.0630 \\
LastFM & 7624 & 27806  & 0.0010 & 7.29 & 216 & 0.0409 \\
PGP		& 10680	& 24316 & 0.0004 & 4.554 & 205 & 0.0559 \\
Sex & 10106	 & 39016 & 0.0008 & 7.721 & 311 & 0.0315 \\
\bottomrule
\end{tabularx}
}
\caption{The basic topological characteristics of 12 real-world networks.}
\label{table:topology}
\end{table}

\subsection{Evaluation Metrics} \label{sec:eval-metric}
To quantitatively evaluate the performance of the 1D-CGS model in ranking influential nodes, we employed three widely used metrics: Kendall’s Tau, Jaccard Similarity, and Monotonicity Relation.

\textbf{Kendall’s Tau ($\tau$) \cite{kendall1938new}.} $\tau$ is a rank correlation metric that measures how well two rankings align. It evaluates the level of agreement by counting the number of concordant and discordant pairs among the ranked items. A pair is considered concordant if both rankings place the items in the same order, and discordant if the order differs. The formula for Kendall’s Tau is:
\begin{equation}
\tau = \frac{C - D}{\frac{1}{2} n(n-1)}
\label{equ:kendall}
\end{equation}
where $C$ is the number of concordant pairs, $D$ is the number of discordant pairs, and $n$ is the total number of ranked nodes. The value of $\tau$ ranges from -1, indicating complete disagreement, to +1, indicating perfect agreement between the two rankings.

\textbf{Jaccard Similarity (JS) \cite{Jacc-1998jaccard}.} JS is a metric used to quantify the similarity between two sets by measuring their overlap. In the context of node ranking, let $A$ and $B$ represent the sets of top-k nodes from the predicted ranking and the ground truth, respectively. The Jaccard Similarity is defined as:
\begin{equation}
JS(A, B) = \frac{|A \cap B|}{|A \cup B|}
\label{equ:Jacc}
\end{equation}
The JS score ranges from 0 to 1, where a higher value indicates a greater overlap between the predicted and actual top-k influential nodes, reflecting better alignment and accuracy of the model’s rankings.

\textbf{Monotonicity Index (MI) \cite{MR-bae2014identifying}.} MI is a metric that evaluates the degree of uniqueness in the predicted ranking list. Specifically, it measures how often the model assigns the same rank to different nodes. A well performing ranking model is expected to give distinct scores to most nodes, allowing for clear differentiation between their influence levels. The MI is calculated as:
\begin{equation}
MI = \left(1-\frac{\sum_{r \in R} N_r \cdot (N_r - 1)}{N_u \cdot (N_u - 1)} \right)^2
\label{equ:MR}
\end{equation}
where $R$ is the set of assigned ranks, $N_r$ is the number of nodes assigned to each rank $r$, and $N_u$
is the total number of unique ranks in the list. The resulting value ranges from 0 to 1, where a score closer to 1 indicates that most nodes have been assigned unique ranks, reflecting a more informative and discriminative ranking.

\subsection{Baseline Methods for Comparison} 
To evaluate the effectiveness of the proposed 1D-CGS model, we compared it against a comprehensive set of baseline methods commonly used for identifying influential nodes in complex networks. These include traditional centrality-based methods such as:

\textbf{Degree Centrality (DC) \cite{DCfreeman-FREEMAN1978215}.} This is one of the simplest centrality measures, which assumes that nodes with more direct connections are more influential. It is computed as the degree of a node normalized by the maximum possible degree:
\begin{equation}
    DC(v_i) = \frac{\deg(v_i)}{n - 1}
\end{equation}
where $\deg(v_i)$ is the number of direct neighbors of node $v_i$, and $n$ is the total number of nodes in the network. A high DC value implies that the node has high influence.

\textbf{Betweenness Centrality (BC) \cite{BC-freeman1977set}.} This measure reflects the extent to which a node lies on paths between other nodes. A node with high betweenness can control or facilitate the flow of information across the network:
\begin{equation}
    BC(v_i) = \sum_{\substack{s \neq v_i \\ \neq t}} \frac{\sigma_{st}(v_i)}{\sigma_{st}}
\end{equation}
where $\sigma_{st}$ is the number of shortest paths from $s$ to $t$, and $\sigma_{st}(v_i)$ is the number of those paths that pass through $(v_i)$.

\textbf{Mixed Degree Decomposition (MDD) \cite{MDD-zeng2013ranking}.} MDD is a hybrid ranking strategy that balances local and global topological information by integrating degree centrality and k-core into a single measure. It introduces a tunable parameter $\lambda \in [0,1]$ to adjust the contribution of each component. Specifically, for each node $v$, the mixed degree $k_m(v)$ is calculated as:
\begin{equation}
    k_m(v) = k_s(v) + \lambda \cdot \bigl( k_d(v) - k_s(v) \bigr)
\end{equation}

where $k_d(v)$ is the node’s original degree in the graph, and $k_s(v)$ is its residual degree during the iterative pruning process. When $\lambda =0$, MDD behaves like standard k-core, and when When $\lambda =1$, it reduces to degree centrality. Nodes are iteratively removed based on their $k_m$ values, and the order in which they are removed forms the final ranking.

\textbf{Neighborhood Degree (ND) \cite{ND-namtirtha2021best}.} ND measures the total degree of a node’s neighbors, indicating how well connected its local neighborhood is. It is defined as:

\begin{equation}
    \text{ND}(v) = \sum_{u \in N(v)} \deg(u)
\end{equation}
where $N(v)$ is the set of neighbors of node $v$, and $\deg(u)$ is the degree of neighbor $u$.

\textbf{H-index (HI) \cite{H-index2016h}.} The H-index, originally introduced to measure the research impact of individuals, has been adapted to network analysis to evaluate node influence. In this context, it reflects how well connected a node's neighbors are. To compute it, the degrees of a node's neighbors are first sorted in descending order. The H-index of a node is then defined as the highest rank $h$ such that the node has at least 
$h$ neighbors with degree $\geq h$. A larger H-index value implies that the node is not only connected to many others but that those neighbors are themselves influential, making it a reliable indicator of a node’s central role in the network.

\textbf{K-core \cite{k-shell2010identification}.} This method identifies a node's importance based on how deeply embedded it is within the network. A node belongs to a 
k-core if it is connected to at least $k$ other nodes in that subgraph. Nodes in higher order cores are considered more topologically central and influential.

\textbf{V-community (Vc) \cite{VC-2008fast}.} Vc estimates a node’s spreading potential based on its ability to bridge different communities within the network. It is computed by counting the number of distinct communities that a node is directly connected to. The idea is that a node linking multiple communities can act as a gateway for information diffusion across otherwise separate regions. Thus, nodes with higher Vc scores are considered more influential.
In addition, we compared against recent deep learning models including RCNN \cite{yu2020rcnn} and MRCNN \cite{ou2022iM-RCNN}, which are designed to capture complex topological patterns in node influence. 

%These baselines offer a diverse representation of heuristic, statistical, and learning-based strategies, enabling a fair and thorough evaluation of the 1D-CGS model across multiple dimensions of performance.

\newcolumntype{P}{>{\centering\arraybackslash}X}
\begin{table}[H] 
\centering
\resizebox{\textwidth}{!}{%
\fontsize{8pt}{9pt}\selectfont
\begin{tabularx}{\textwidth}{lCCCCCCCCCC}
\toprule
\textbf{Net.}	& \textbf{BC}	& \textbf{DC} & \textbf{HI} & \textbf{K-core} & \textbf{Vc} &  
 \textbf{MDD} & \textbf{ND} & \textbf{MRCNN} &   \textbf{RCNN} &  
 \textbf{\mbox{1D-CGS}}\\
\midrule
Jazz  & 0.4733 & 0.8268  & 0.8268 & 0.7787 & 0.2748 & 0.8607 & 0.9211 & 0.8126 & 0.8435 & \textbf{0.9280} \\

%CEnew  & 0.5135 & 0.7015 & 0.7354   & 0.7453 & 0.5705 & 0.6606 & 0.7977 & \textbf{0.8579} & 0.8461 & 0.8245 \\

Email& 0.6460 & 0.8106  & 0.8393  & 0.8252 & 0.7556 &  0.8048 & 0.9110 & 0.9134 & 0.8467 & \textbf{0.9219} \\

%Faa & 0.2634 & 0.4995  & 0.5779   & 0.5757 & 0.5723 &  0.4826 & 0.6907 & 0.7282 & 0.6390 & \textbf{0.7449} \\

%NS  & 0.3516 & 0.6507 & 0.6309  & 0.6295 & 0.3525 &   0.6060 & \textbf{0.8175} & 0.6790 & 0.5953 & 0.7556 \\

Stelzl  & 0.4065 & 0.4630 & 0.4919 & 0.4910 & 0.5772 & 0.4786 & 0.7299 & 0.6885 & 0.6672 & \textbf{0.7617} \\

Figeys  &  0.4397 & 0.4767  & 0.4989 & 0.4984 & 0.4835 &  0.5387 & 0.6342 & \textbf{0.6632} & 0.5876 & 0.6337 \\

Hamster	& 0.4945 & 0.8203  & 0.6962   & 0.7001 & 0.7001 &  0.7140 & 0.8366 & 0.8299 & 0.8142 & \textbf{0.8379} \\

%Vidal  & 0.5390 & 0.6003  & 0.6442 & 0.6445 & 0.6803 &  0.5322 & 0.7649 & 0.7667 & 0.7188 & \textbf{0.7853} \\

%Citeseer 	& -	 & -  & -   & - & - & - & - & - & - \\

Facebook & 0.4250 & 0.6310 & 0.6671  & 0.6678 & 0.4126 &  0.6487 & 0.7380 & \textbf{0.7759} & 0.7244 & 0.7167\\

EPA	& 0.4598 & 0.4918  & 0.4953   & 0.4939 & 0.5269 &  0.4785 & 0.6531 & \textbf{0.6925} & 0.6832 &  0.6812\\

%Power	& -	 & -  & -   & - & - & -  & - & - & - \\

Router	& 0.1770 & 0.1810  & 0.0779  & 0.0776 & 0.4117 &  0.3251 & 0.6090 & 0.4880 & 0.6126 & \textbf{0.7104} \\

GrQ  & 0.4227 & 0.6271 & 0.6264  & 0.6237 & 0.5666 &  0.5936 & 0.7529 & 0.6911 & 0.6848 & \textbf{0.7614} \\

%Delaunay & -	 & -  & -   & - & - & - & - & - & - \\
LastFM & 0.3801 & 0.5454  & 0.5749 & 0.5764 & 0.5257 & 0.5589 & 0.7115 & 0.7126 & 0.6507 & \textbf{0.7381} \\

%Hep & 0.5180 & 0.5892  & 0.6217 & 0.6184 & 0.7038 &  0.5501 & 0.7590 & 0.7890 & 0.7372 & \textbf{0.7913} \\

PGP	& 0.2794 & 0.4065  & 0.3998   & 0.3989 & 0.5341 &  0.4188 & 0.6406 & 0.6140 & 0.6029 &  \textbf{0.7281} \\

Sex & 0.5623 & 0.5989  & 0.6333   & 0.6338 & 0.6337 &  0.6072 & 0.8208 & 0.8065 & 0.7464 & \textbf{0.8364} \\

\bottomrule
\end{tabularx}
}
\caption{Average Kendall’s Tau correlation scores across all networks for the proposed 1D-CGS model and baseline methods.}
\label{table:kendall}
\end{table}

\begin{figure}[H]
    \centering
    \begin{tikzpicture}
        \node (img1) at (0, 0) {\includegraphics[width=0.95 \textwidth]{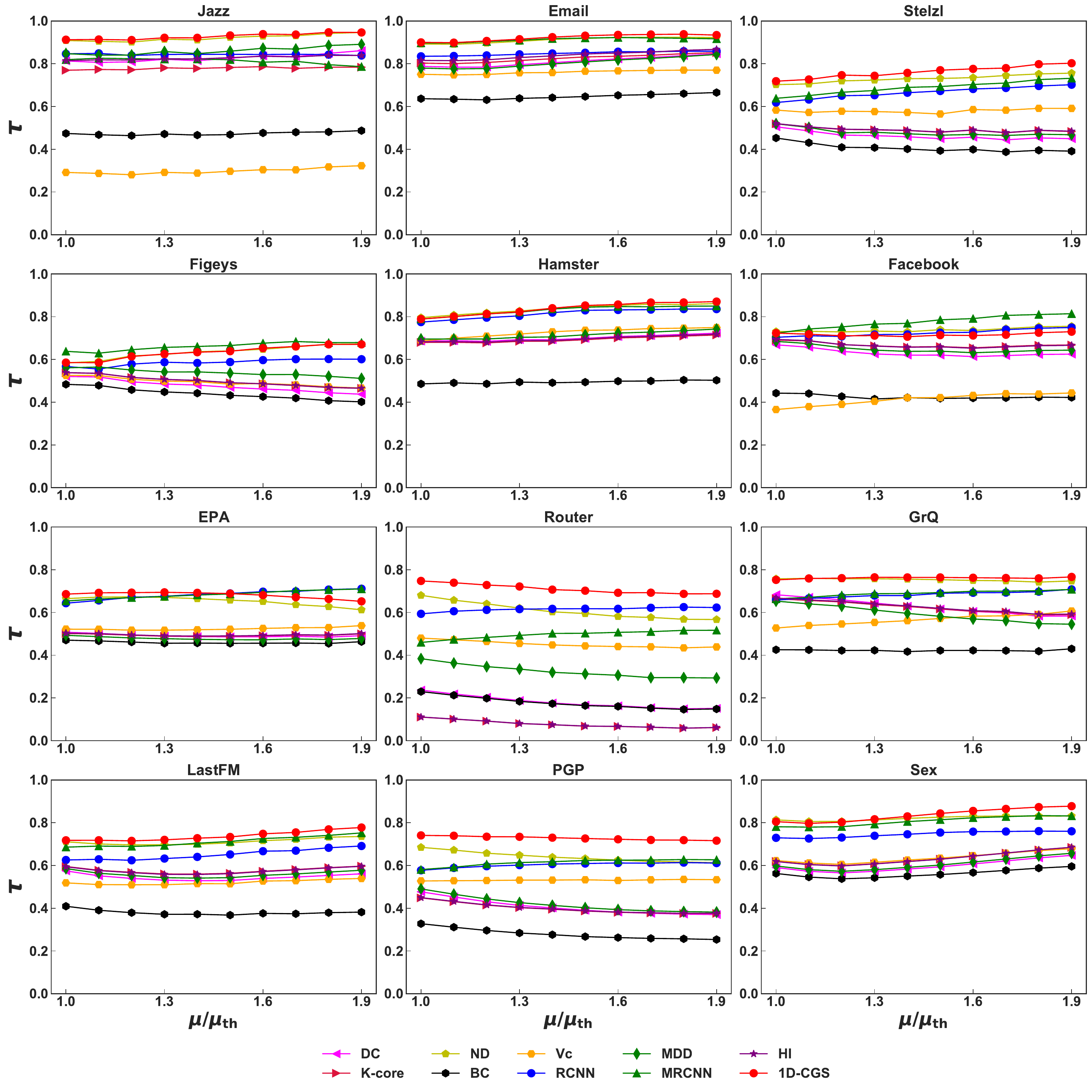}};
    \end{tikzpicture}
    \caption{Kendall’s $\tau$ correlation between model generated rankings and ground-truth rankings obtained from the SIR model, evaluated across varying infection rates ($\mu/\mu_{\mathrm{th}}$) }
\label{img:kendall}
\end{figure}

\subsection{Results}\label{sec:results}
This section presents and analyzes the performance of the proposed 1D-CGS model compared to baseline methods. We evaluate ranking quality using $\tau$, JS, and MI across a wide range of real world network datasets.

\textbf{Ranking accuracy via $\tau$.} As shown in Table \ref{table:kendall} and illustrated in Figure \ref{img:kendall}, the proposed 1D-CGS model consistently achieves the highest Kendall’s $\tau$ scores across almost all tested networks. These scores quantify the correlation between the predicted rankings and the ground truth node rankings derived from epidemic simulations using the SIR model. While Table \ref{table:kendall} presents the average $\tau$, Figure \ref{img:kendall} visualizes the model's robustness across a range of infection rates, offering further insight into its stability under varying diffusion conditions. Compared to baselines, including RCNN and MRCNN, 1D-CGS consistently exhibits superior performance. In terms of relative improvement over RCNN, the most substantial gains are observed in networks such as PGP 12.52\%, Stelzl 9.45\%, and Sex 9.0 \%. Similarly, the model outperforms MRCNN by 11.41\%, 7.32\%, and 2.99\% in these networks, respectively. Even in cases where MRCNN performs competitively—such as EPA 1.13\% gap and Figeys 2.95\%, 1D-CGS maintains a consistent lead.
A particularly striking result is seen in the Router network, where 1D-CGS achieves a $\tau$ score of 0.7104, corresponding to a 22.24\% improvement over MRCNN and 9.78\% over RCNN. This performance underscores the model’s ability to handle sparse (less density) networks where other methods struggle.
Additional performance gaps appear in networks such as Jazz, GrQ, and LastFM. For instance, in Jazz, 1D-CGS improves upon RCNN and MRCNN by 8.45\% and 11.54\%, respectively. In both GrQ and LastFM, the margins exceed 7\% over RCNN, demonstrating that 1D-CGS is consistently more aligned with ground truth rankings, regardless of the network's internal connectivity or structure.
While ND occasionally ranks closely behind 1D-CGS and even outperforms some deep models in some networks, it lacks the adaptability to generalize across topologies. Its strong performance reinforces the importance of neighborhood information, but its fixed, non learnable formulation limits its effectiveness in comparison to the learning-based flexibility of 1D-CGS.

\begin{comment}
In the Jazz network, 1D-CGS reaches a $\tau$ score of 0.9280, outperforming all other methods. The second-best method here is ND (0.9211). While ND performs closely, 1D-CGS provides better differentiation by leveraging both local features through CNNs and relational context through GraphSAGE. Similarly, in Email, 1D-CGS records a $\tau$ of 0.9219, again outperforming MRCNN (0.9134) and RCNN (0.8467) by clear margins. These results highlight 1D-CGS’s ability to capture influence patterns more effectively, even when other deep models perform competitively.
In Stelzl, 1D-CGS achieves 0.7617, exceeding ND (0.7299) by 3.18\%. MRCNN and RCNN reach only 0.6885 and 0.6672, respectively. This consistent gap suggests that while deep learning models can capture complex structures, their performance may degrade in settings with sparser topology—conditions where 1D-CGS remains robust. A similar pattern appears in Figeys, where 1D-CGS scores 0.6337, marginally behind ND (0.6342) by only 0.05\%. However, 1D-CGS still significantly outperforms MRCNN (0.6632) and RCNN (0.5876) in overall ranking quality.
\end{comment}

\newcolumntype{D}{>{\centering\arraybackslash}X}
\begin{table}[H] 
\centering
\resizebox{\textwidth}{!}{%
\fontsize{8pt}{9pt}\selectfont
\begin{tabularx}{\textwidth}{lCCCCCCCCCC}
\toprule
\textbf{Net.}	& \textbf{BC}	& \textbf{DC} & \textbf{HI} & \textbf{K-core} & \textbf{Vc} &  
 \textbf{MDD} & \textbf{ND} & \textbf{MRCNN} & \textbf{RCNN} &  
 \textbf{\mbox{1D-CGS}}\\
\midrule

Jazz  & 0.7378 & 0.8896  & 0.8587 & 0.8410 & 0.6733 & 0.9071 & 0.9443 & 0.8588 & 0.9021 & \textbf{0.9509} \\

%CEnew  & 0.5502 & 0.7374 & 0.7408 & 0.7354 & 0.6681 &  0.7342 & 0.7927 & 0.8273 & 0.8068 & 0.8280\\

Email	& 0.4883 & 0.7129  & 0.7140 & 0.5774 & 0.4955 & 0.7509 & \textbf{0.8641}& 0.7468 & 0.7643 & 0.8611 \\

%Faa  & 0.2621 & 0.4754  & 0.5907  & 0.5470 & 0.4437 & 0.5003 & 0.6430 & \textbf{0.6947} & 0.5752 & 0.6653 \\

%NS  & 0.2795 & 0.3359 & 0.2293 & 0.2246 & 0.3056 & 0.2474 & 0.3150 & 0.4117 & \textbf{0.4338} & 0.3873 \\

Stelzl & 0.3232 & 0.4463  & 0.6864 & 0.6427 & 0.5543 &  0.5027 & 0.6910 & 0.7548 & 0.6861 & \textbf{0.7673} \\

Figeys & 0.3373 & 0.4263  & 0.5013 & 0.4827 &  0.3435 &  0.4636 & 0.4274 & 0.5329 & 0.3620 & \textbf{0.5506} \\

Hamster	& 0.3671 & 0.6321  & 0.6321 & 0.4978 & 0.4096 &  0.5428 & 0.7695 & 0.5879 & 0.6463 & \textbf{0.7729} \\

%Vidal & 0.4020 & 0.4801  & 0.5915 & 0.5138 & 0.5168 &  0.5082 & 0.6008 & \textbf{0.6849} & 0.6383 & 0.6702 \\

%Citeseer 	& -	 & -  & -   & - & - & - & - & - & - \\

Facebook & 0.0398 & 0.4398  & 0.6325  & 0.5609 & 0.0279 & 0.5373 &  0.6081 & 0.4539 & 0.4707 & \textbf{0.6379} \\

EPA	& 0.4933 &  0.5644 & 0.5038 & 0.3655 & 0.5368 &  0.5991 & 0.5071 & 0.6230 & 0.5809 & \textbf{0.7026} \\

%Power	& -	 & -  & -   & - & - & -  & - & - & - \\

Router	& 0.3979	 & 0.4380  & 0.5438 & 0.5263 & 0.5263 & 0.4678 & 0.6827 & \textbf{0.7524} & 0.6765 & 0.6677 \\

GrQ & 0.0941 & 0.4217  & 0.3994 & 0.3929 & 0.1120 &  0.4217 & 0.4663 & 0.5159 & \textbf{0.5304} & 0.4739 \\

%Delaunay & -	 & -  & -   & - & - & - & - & - & - \\

LastFM  & 0.1686 & 0.3829 & 0.5778 & 0.5281 & 0.1728 & 0.4209 & 0.6155  & 0.5800 & 0.6185 & \textbf{0.6430} \\

%Hep & 0.2184 & 0.4030  & 0.2383  & 0.1614 & 0.3342 &  0.3534 & 0.4182 & 0.2820 & \textbf{0.5545} & 0.4070 \\

PGP	& 0.0976 & 0.4835  & 0.6356 & 0.6267 & 0.2431 &  0.5710 & 0.7918 & 0.6863 & 0.7068 & \textbf{0.8026}\\

Sex  & 0.4785 & 0.5177 & 0.6992 & 0.3578 & 0.3453 & 0.5596 & 0.5441 & 0.5864 & 0.3052 & \textbf{0.7690} \\

\bottomrule
\end{tabularx}
}
\caption{ Average results of JS generated by 1D-CGS model and different methods.}
\label{table:Jacc}
\end{table}

\begin{figure}[H]
    \centering
    \begin{tikzpicture}
        \node (img1) at (0, 0) {\includegraphics[width=0.95 \textwidth]{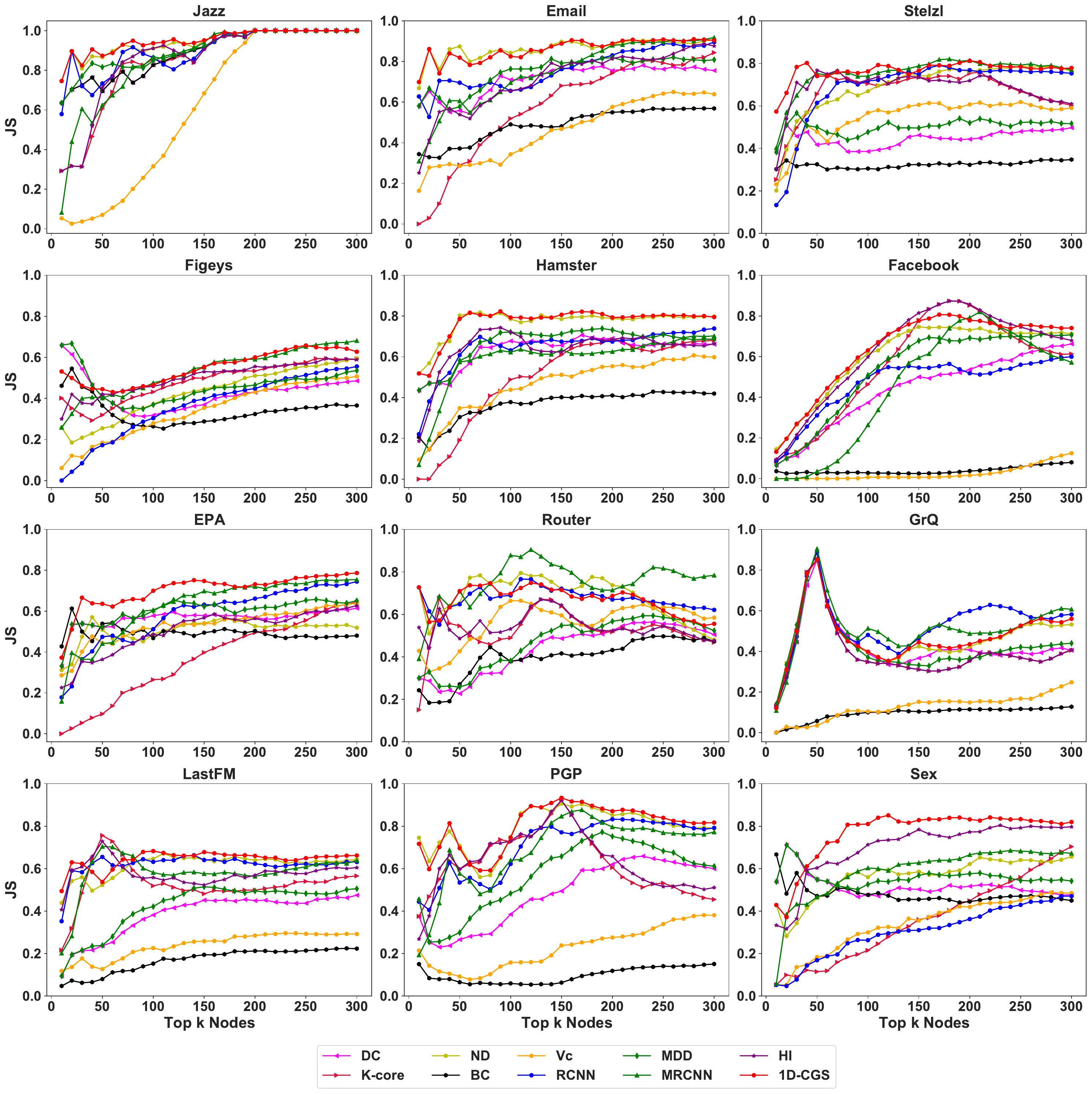}};
    \end{tikzpicture}
    \caption{Jaccard Similarity between top-k nodes from various methods and ground truth generated from SIR model.}
\label{img:jacc}
\end{figure}

\textbf{Top-ranked influential nodes.} To further assess the accuracy of node rankings, we evaluate the overlap between the top-ranked influential nodes predicted by each method and those identified by the SIR based ground truth using JS \ref{equ:Jacc}. Table \ref{table:Jacc} presents the average JS scores across the tested networks, while Figure \ref{img:jacc} visualizes the overlap scores across different k-values, highlighting the consistency of each method.
The proposed model achieves highest JS scores than baselines in the majority of networks. For example, in the Jazz network, 1D-CGS improves JS by 9.21\% over MRCNN and 4.88\% over RCNN. In Hamster, it achieves a 12.66\% gain over RCNN and a 18.5\% improvement over MRCNN. In the Facebook network, JS improves by 16.72\% and 18.4\% relative to RCNN and MRCNN, respectively. The most significant improvement is observed in the Sex network, where 1D-CGS achieves a 46.38\% increase over RCNN and a 18.4\% gain over MRCNN. 
On the other hand, in GrQ network, RCNN and MRCNN slightly outperform 1D-CGS. Although ND performs reasonably well in some networks, but these methods struggle to generalize in other networks.
Furthermore, as illustrated in Figure \ref{img:jacc}, generally the proposed model achieves the highest JS in the top 10 ranked nodes across all network datasets.

\begin{figure}[H]
    \centering
    \begin{tikzpicture}
        \node (img1) at (0, 0) {\includegraphics[width=0.95 \textwidth]{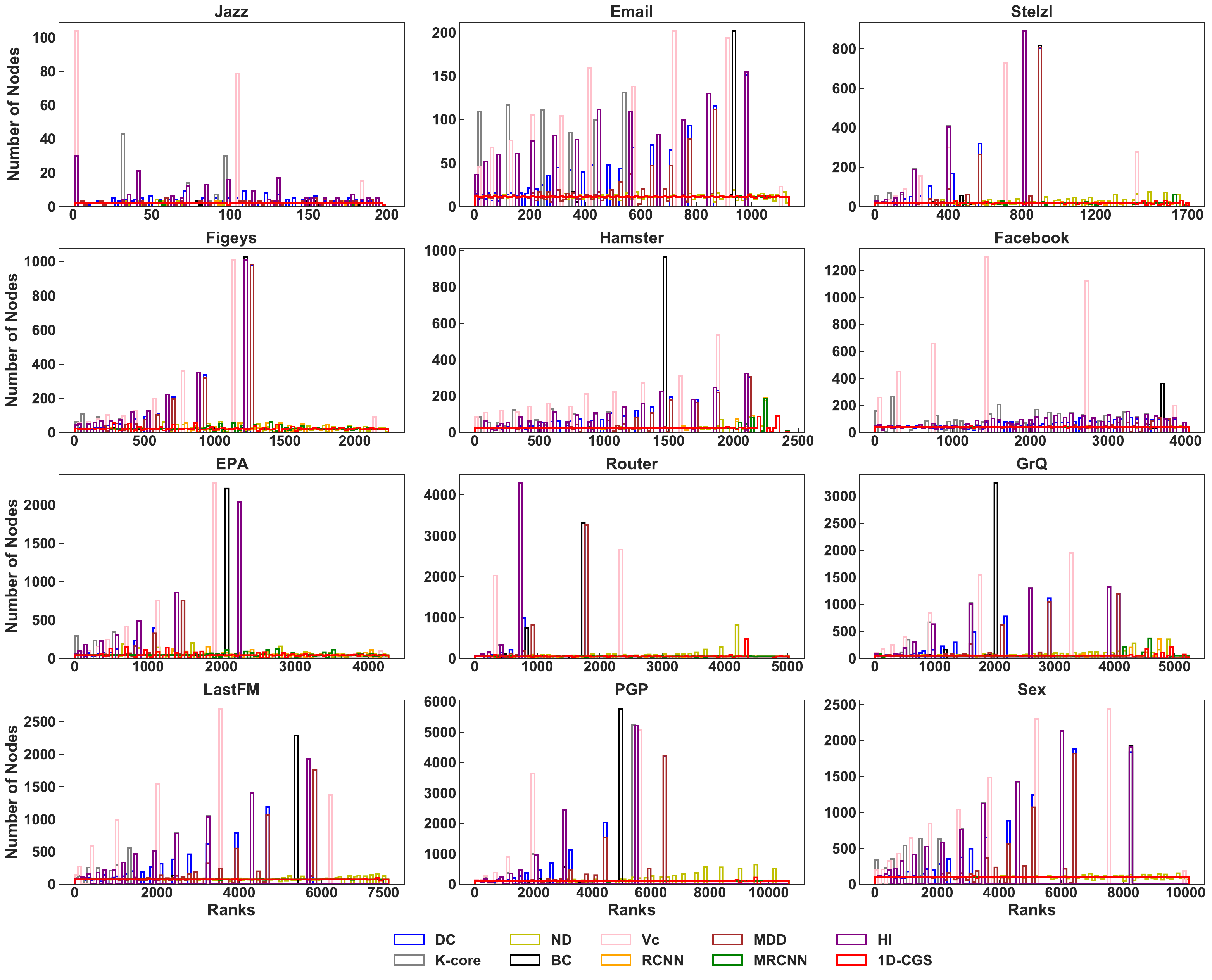}};
    \end{tikzpicture}
    \caption{Histograms illustrating the distribution of node ranks assigned by each method across twelve real-world networks.}
\label{img:Ranking-Distribution}
\end{figure}

\textbf{The uniqueness score of ranking.} Table \ref{table:mr} demonstrates that the 1D-CGS model achieves exceptional ranking distinctiveness, with near perfect MI scores in Jazz (1.0), Email (0.9999), and Facebook (0.9999) networks. Both RCNN and MRCNN also show strong performance with high MI values, confirming the effectiveness of deep learning approaches for this task. However, 1D-CGS maintains consistent discriminative capability across all tested networks. While other traditional centrality measures exhibit significantly lower MI scores overall, ND emerges as a competitive baseline, though still less consistent than 1D-CGS. 
In support of the MI results presented in Table \ref{table:mr}, Figure \ref{img:Ranking-Distribution} displays rank distribution histograms that illustrate how each method distributes influence scores across nodes in various network datasets. The three deep learning-based models RCNN, MRCNN, and 1D-CGS all achieve excellent ranking uniqueness, as reflected in their high MI scores and well spread histograms. These results confirm their capacity to assign distinct scores to most nodes, which is essential for producing meaningful and well separated rankings. Among them, 1D-CGS consistently exhibits an almost perfect distribution of ranks, indicating highly refined discriminative power across networks of varying size and density. While RCNN and MRCNN also produce highly distinctive rankings, their histograms at times show minor rank duplication not observed in 1D-CGS. In contrast, traditional centrality-based methods tend to assign many identical ranks, resulting in less discriminative distributions. Overall, the rank histograms in Figure \ref{img:Ranking-Distribution}, together with the MI values in Table \ref{table:mr}, confirm that 1D-CGS is highly effective in producing unique and consistent rankings across all tested networks.

\newcolumntype{R}{>{\centering\arraybackslash}X}
\begin{table}[H] 
\centering
\resizebox{\textwidth}{!}{%
\fontsize{8pt}{9pt}\selectfont
\begin{tabularx}{\textwidth}{lCCCCCCCCCC}
\toprule
\textbf{Net.}	& \textbf{BC}	& \textbf{DC} & \textbf{HI} & \textbf{K-core} & \textbf{Vc} &  
 \textbf{MDD} & \textbf{ND} & \textbf{MRCNN} & \textbf{RCNN} &  
 \textbf{\mbox{1D-CGS}}\\
\midrule
Jazz  & 0.9885 & 0.9659 & 0.9659  & 0.7944 & 0.2086 &  0.9919 & 0.9980 & 0.9993 & 0.9991 & \textbf{1.0} \\

%CEnew  & 0.8741	 & 0.7921 & 0.7310 & 0.6961 & 0.6974 &  0.8755 & 0.9936 & 0 & 0.9975 & \textbf{0.9988} \\

Email & 0.9400 & 0.8873 & 0.8582 &  0.8088 & 0.7701 & 0.9227 & 0.9940 & 0.9998 & 0.9998 & \textbf{0.9999} \\

%Faa  & 0.9770 & 0.5921  & 0.4641 &  0.3771 & 0.5662 & 0.6608 & 0.9609 & \textbf{0.9998} & \textbf{0.9998} & 0.9995 \\

%NS  &  0.1049	 & 0.7069 & 0.6710 & 0.6633 &  0.0492 & 0.7054  &  0.7419 & 0.8965 & 0.9171 & \textbf{0.9682 } \\

Stelzl  & 0.5995 & 0.5287  & 0.4509  & 0.4260 & 0.5479 & 0.5373 & 0.9682 & 0.9956 & 0.9956 & \textbf{0.9963}\\

Figeys  & 0.6233 & 0.5927 & 0.5708 &  0.5617 & 0.5757 &  0.6058 & 0.9887 & \textbf{0.9938} & \textbf{0.9938} & \textbf{0.9938} \\

Hamster	& 0.7123 & 0.8979  & 0.8835 & 0.8714 & 0.7835 &  0.9268 & 0.9823 & 0.9856 & 0.9856 & \textbf{0.9937}  \\

%Vidal  & 0.6656	 & 0.6507 & 0.5865  & 0.5569& 0.6238 &  0.6666 & 0.9712 & 0.9923 & 0.9923 & \textbf{0.9945} \\
%Citeseer 	& -	 & -  & -   & - & - & - & - & - & - \\

Facebook & 0.9855 & 0.9739 & 0.9664 & 0.9419 & 0.5979 & 0.9914 & 0.9995 & 0.9998 & 0.9998  & \textbf{0.9999}\\

EPA	& 0.5344 & 0.5285 & 0.5080 & 0.4940 & 0.4512 &  0.5393 & 0.9786 & \textbf{0.9895} & \textbf{0.9895} & \textbf{0.9895} \\

%Power	& -	 & -  & -   & - & - & -  & - & - & - \\

Router	& 0.3043 & 0.2886 & 0.0875 &  0.0691 & 0.3023 & 0.3030 & 0.9199 & \textbf{0.9968} & 0.9966 & 0.9807 \\

GrQ  & 0.3822 & 0.7460 & 0.6906 & 0.6631 & 0.5460 & 0.7935 & 0.9649 & 0.9872 & 0.9871 & \textbf{0.9942} \\

%Delaunay & -	 & -  & -   & - & - & - & - & - & - \\
LastFM & 0.8346 & 0.7977 & 0.7570 & 0.7570 & 0.5976 &  0.8400 & 0.9886 & \textbf{0.9998} & \textbf{0.9998}  & 0.9997 \\

%Hep & 0.5077 & 0.7627  & 0.7070 & 0.6742 & 0.6407 &  0.8092 & 0.9736 & 0 & 0.9935 & \textbf{0.9969} \\

PGP	& 0.5120 & 0.6193  & 0.5172 & 0.4805 & 0.4220 & 0.6651 & 0.9551 & \textbf{0.9997} & \textbf{0.9997} & 0.9985  \\

Sex & 0.9336 & 0.8057  & 0.7725 & 0.7530 & 0.7131 &  0.8419 & 0.9937 & \textbf{0.9999} & \textbf{0.9999}  & \textbf{0.9999} \\
\bottomrule
\end{tabularx}
}
\caption{MI scores for each method across all networks.}
\label{table:mr}
\end{table}

\subsection{Practical Running Time}\label{sec:RT}
\begin{figure}[H]
    \centering
    \begin{tikzpicture}
        \node (img1) at (0, 0) {\includegraphics[width=0.65 \textwidth]{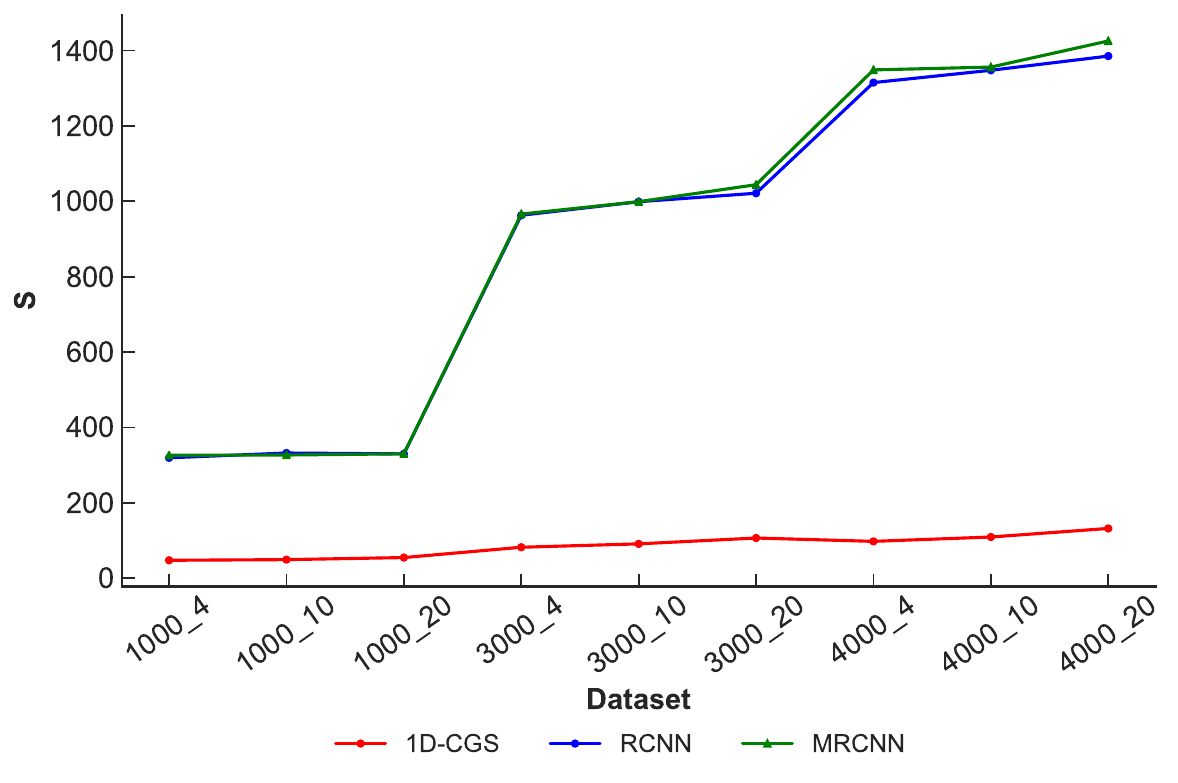}};
    \end{tikzpicture}
    \caption{Training time comparison (in seconds) of 1D-CGS, RCNN, and MRCNN models on synthetic BA networks of varying sizes and average degrees.}
\label{img:RT-T}
\end{figure}

In addition to evaluating predictive accuracy, we analyze the computational efficiency of the proposed model from both empirical and theoretical perspectives. Efficiency is a crucial factor when deploying models in large scale or real time scenarios. Figure \ref{img:RT-T} presents the training times of 1D-CGS, RCNN, and MRCNN models on synthetic BA networks ranging with size of 1000, 3000, and 4000 nodes and average degrees 4, 10, and 20. 

According to prior work, both RCNN and MRCNN adopt a similar architectural foundation based on 2D-CNN. However, they differ in how they construct their input representations. RCNN uses a single channel input matrix per node, with each node represented by a fixed size matrix . The computational complexity of training RCNN is primarily influenced by the number of iterations and the convolutional operations. Specifically, if $D$ denotes the number of convolutional layers, $M_d$ size of the feature matrices at layer $d$, $K_d$ is the convolutional kernel size, and $C_{d-1}$ and $C_{d}$ are the numbers of input and output channels at layer $d$, then the total training complexity of RCNN is given by:
\begin{equation}
    O\left(n + I\cdot \sum_{d=1}^D M_d^2 K_d^2 C_{d-1} C_d \right)
\end{equation}
where $n$ is represent preprocessing time to compute degree of node where the input matrix is based on it, and $I$ is number of training iterations.

MRCNN extends RCNN by incorporating a richer topological representation. It constructs a three-channel matrix for each node using ND, Vc, and k-core as input features. Together, the input construction step of MRCNN has an overall complexity of 
$O(m + n \log n)$. 
)
Since MRCNN retains the same 2D-CNN structure as RCNN, its overall training complexity becomes:

\begin{equation}
    O\left(m + n \log n + I\cdot\sum_{d=1}^{D} M_d^2 K_d^2 C_{d-1} C_d \right)
\end{equation}

In contrast to RCNN and MRCNN, the 1D-CGS model is designed with a lightweight architecture that simplifies both the input representation and model computation. Instead of constructing subgraph based matrices or using multi channel inputs, each node is described by a compact 2-dimensional feature vector, consisting of the node's degree and average neighbor degree, which computed in linear time $O(n + m)$
. These features are processed by two 1D convolutional layers, followed by two GraphSAGE layers for neighborhood aggregation.
the total training complexity of 1D-CGS is approximated by:
\begin{equation}
    O\left(n + m + I \cdot \sum_{d=1}^{D} M_d K_d C_{d-1} C_d + Tml\right)
\end{equation}
where $Tml$ is GraphSAGE complexity with $T$ layers, aggregating over edges. This architecture avoids subgraph matrix construction and reduces both input size and convolutional cost. Empirically, this theoretical advantage is reflected in training and testing times. As shown in Figure \ref{img:RT-T}, 1D-CGS consistently achieves the shortest training times among deep models, even when input construction is included. On synthetic BA networks with increasing size and connectivity, 1D-CGS remains an order of magnitude faster than both RCNN and MRCNN. 

Similarly, Table \ref{table:runTime} shows that during testing, 1D-CGS outperforms all baselines including centrality-based methods and deep models on 12 real world networks. 
Notably, in the Sex network, 1D-CGS requare just 0.6246 seconds. In contrast, RCNN and MRCNN require 102.0722 and 106.0286 seconds, respectively. A similar trend is observed in the PGP network, where 1D-CGS requires only 0.2042 seconds, while RCNN and MRCNN take over 71 and 74 seconds, respectively.
On smaller networks, the performance gap remains consistent. For example in Jazz network, 1D-CGS completes inference in 0.0010 seconds, significantly faster than RCNN (1.10s) and MRCNN (1.22s), and also outperforming other methods such as MDD (0.0658s), K-core (0.0062s), HI (0.0131s), BC (0.9037s). 

In this study, all experiments were conducted using Python 3.8.3 in a 64-bit Windows environment on a machine equipped with an Intel Core i5-8250U CPU and 8 GB RAM.

\newcolumntype{K}{>{\centering\arraybackslash}X}
\begin{table}[H] 
\centering
\resizebox{\textwidth}{!}{%
\fontsize{8pt}{9pt}\selectfont
\begin{tabularx}{\textwidth}{lCCCCCCCCCC}
\toprule
\textbf{Net.}	& \textbf{BC}	& \textbf{DC} & \textbf{HI} & \textbf{K-core} & \textbf{Vc} &  
 \textbf{MDD} & \textbf{ND} & \textbf{MRCNN} & \textbf{RCNN} &  
 \textbf{\mbox{1D-CGS}}\\
\midrule

Jazz  & 0.9037 & 0.0001 & 0.0131 & 0.0062 & 0.0097 & 0.0658 & 0.0008 & 2.3589 & 2.3423 & 0.0010 \\

%CEnew  & -	& -  & - & - & - & - & - & - & - & - \\

Email	& 17.0770 & 0.0028 & 0.0286 & 0.0211 & 0.0091 & 0.1895 & 0.0011 & 9.2719 & 9.2180 & 0.0323 \\

Stelzl & 25.7447 & 0.0040 & 0.0200 & 0.0394 & 0.0025 & 0.1557 & 0.0131 & 9.6238 & 8.6921 & 0.0323 \\

Figeys & 60.3307 & 0.0011 & 0.0323 & 0.0323 & 0.0010 & 0.2041 & 0.0011 & 21.1721 & 20.5931 & 0.0479 \\

Hamster	& 71.3318 & 0.0030 & 0.0957 & 0.0701 & 0.0227 & 0.5903 & 0.0163 & 26.1263 & 25.2105 & 0.0640 \\

%Vidal & 86.3120 & 0.0089 & 0.0589 & 0.0541 & 0.0225 & 0.2943 & 0.0012 & 16.7266 & 15.9143 & 0.0433 \\

%Citeseer & -	& -  & - & - & - & - & - & - & - & - \\

Facebook & 420.9879 & 0.0097 & 0.4790 & 0.5458 & 0.0656 & 4.0670 & 0.0638 & 90.4139 & 79.4247 & 0.3157 \\

EPA	& 195.7634 & 0.0090 & 0.1067 & 0.0717 & 0.0318 & 0.4180 & 0.0306 & 25.2340 & 23.6954 & 0.0792 \\

%Power	& -	& -  & - & - & - & - & - & - & - & - \\

Router	& 258.6012	& 0.0010 & 0.0792 & 0.0479 & 0.0239 & 0.2747 & 0.0170 & 25.0435 & 22.5337 & 0.0792 \\

GrQ & 252.8678 & 0.0100 & 0.1224 & 0.0635 & 0.0323 & 0.8305 & 0.0214 & 27.0937 & 25.3964 & 0.1094 \\

%Delaunay & -	& -  & - & - & - & - & - & - & - & - \\

LastFM & 868.3343 & 0.0036 & 0.1898 & 0.1549 & 0.0354 & 1.0926 & 0.0448 & 62.6705 & 58.0663 & 0.1573 \\

%Hep & -	& -  & - & - & - & - & - & - & - & - \\

PGP	& 1475.0845	& 0.0178 & 0.2198 & 0.1565 & 0.0781 & 1.3850 & 0.0571 & 74.9367 & 71.3503 & 0.2042 \\

Sex & 1772.8054	& 0.0130 & 0.5523 & 0.2714 & 0.0820 & 1.7377 & 0.0710 & 106.0286 & 102.0722 & 0.6246 \\

\bottomrule
\end{tabularx}
}
\caption{Testing time in seconds across all networks for the proposed 1D-CGS model and baselines.}
\label{table:runTime}
\end{table}

\section{Conclusion} \label{sec:conclusion}
Identifying influential nodes in complex networks is a fundamental problem with wide ranging applications, from epidemic control to spread information efficiently. While many existing studies have made notable contributions, they often suffer from limited accuracy or high computational costs.
To address these limitations, we proposed 1D-CGS, a hybrid model that combines the efficiency of 1D-CNN with the topological learning capabilities of GraphSAGE. This design allows the model to extract meaningful representations from two simple yet powerful features: node degree and average neighbor degree. The model maintains a lightweight architecture while achieving strong learning ability.
Extensive experiments on twelve real world networks demonstrated that 1D-CGS consistently outperforms centrality-based methods and recent DL-based methods. Notably, the model achieves significant improvements in both Kendall’s Tau and Jaccard Similarity, indicating better alignment with ground truth influence rankings obtained via SIR simulations. Furthermore, 1D-CGS offers near perfect rank uniqueness and significantly reduced prediction times. These results highlight the suitability of 1D-CGS for large scale networks and scenarios requiring efficient computation.
These findings highlight the effectiveness of the proposed hybrid model and offer valuable insights for designing accurate and efficient node ranking algorithms in real world networks.
In future work, we intend to extend and refine our model to operate effectively on weighted networks, enabling more accurate representation of real world networks.

\bibliographystyle{elsarticle-num} % Use Elsevier numeric style
\bibliography{article}

\end{document}